\documentclass[
 reprint,
 amsmath,amssymb,
 aps,
 prl,
 superscriptaddress
]{revtex4-2}

\usepackage[utf8]{inputenc}
\usepackage{graphicx}
\usepackage{hyperref}
\usepackage{siunitx}
\usepackage{gensymb}
\usepackage{textcomp}
\usepackage[version=4]{mhchem}
\newcommand{\figsize}{1.}

\begin{document}

\preprint{APS/123-QED}

\title{Radium Ion Optical Clock}

\author{C. A. Holliman}
\affiliation{Department of Physics, University of California, Santa Barbara, California 93106, USA}
\email{jayich@gmail.com}
\author{M. Fan}
\affiliation{Department of Physics, University of California, Santa Barbara, California 93106, USA}
\author{A. Contractor}
\affiliation{Department of Physics, University of California, Santa Barbara, California 93106, USA}
\author{S. M. Brewer}
\affiliation{Department of Physics, Colorado State University, Fort Collins, Colorado 80523, USA}
\author{A. M. Jayich}
\affiliation{Department of Physics, University of California, Santa Barbara, California 93106, USA}

\date{\today}

\begin{abstract}
We report the first operation of a Ra$^{+}$ optical clock, a promising high-performance clock candidate.  The clock uses a single trapped $^{226}$Ra$^{+}$ ion and operates on the $7s\ ^2S_{1/2}\rightarrow$ $6d\ ^2D_{5/2}$ electric quadrupole transition.  By self-referencing three pairs of symmetric Zeeman transitions, we demonstrate a frequency instability of 1.1$\times10^{-13}$/$\sqrt{\tau}$, where $\tau$ is the averaging time in seconds.  The total systematic uncertainty is evaluated to be ${\Delta \nu / \nu = 9 \times 10^{-16}}$.  Using the clock, we realize the first measurement of the ratio of the $D_{5/2}$ state to the $S_{1/2}$ state Land\'{e} $g$-factors: $g_{D}/g_{S}$ = \SI{0.5988053\pm.0000011}.  A Ra$^{+}$ optical clock could improve limits on the time variation of the fine structure constant, $\dot \alpha / \alpha$, in an optical frequency comparison. The ion also has several features that make it a suitable system for a transportable optical clock.

\end{abstract}

\maketitle
Optical clocks, based on narrow-linewidth atomic transitions, are the most precise instruments ever realized \cite{BACON2021Nature}.  The performance of several optical clocks, using different atoms, have now surpassed that of the primary cesium frequency standard~\cite{Brewer2019,Huntemann2016,Nicholson2015,McGrew2018}, which marks a significant advance toward a proposed redefinition of the second~\cite{Riehle2015}.  Optical clocks also have the potential to uncover new physics beyond the standard model at the high-precision, low-energy  frontier, including searches for ultralight scalar dark matter \cite{Derevianko2014}, the time variation of fundamental constants \cite{Safronova2018RMP, Lange2021}, and violations of Einstein's equivalence principle \cite{Takamoto2020}.  In an effort to improve clock performance, atomic systems that are less sensitive to limiting systematic uncertainties such as Lu$^{+}$ \cite{Arnold2018} and Ba$^{4+}$ \cite{Beloy2020} have been proposed.  For both advancing clock performance and for their enhanced sensitivity to potential sources of new physics, systems including highly-charged ions \cite{Kozlov2018} and a radioactive thorium nuclear clock \cite{Flambaum2006, Seiferle2019} are being pursued.

The radium ion is well suited to realizing a high-performance optical clock \cite{NunezPortela2014}.  The $7s\ ^2S_{1/2}\rightarrow$ $6d\ ^2D_{5/2}$ electric quadrupole clock transition ($\lambda = 728$~nm, $\tau\approx300$~ms, $\Gamma / 2 \pi = 0.5$~Hz) in Ra$^{+}$ has a small, negative differential static scalar polarizability (DSSP) $\Delta \alpha_0 = -22.2(1.7)$~a.u.~\cite{Sahoo2009a}.  This leads to both a small frequency shift due to the blackbody radiation (BBR) environment and allows for clock operation at a trap drive frequency (6.2~MHz) such that the micromotion-induced scalar Stark shift and the second-order Doppler shift cancel \cite{Dube2014, Huang2019}.  Along with the expected clock performance, the radium ion has the largest positive enhancement to the time variation of the fine structure constant, $\kappa_{\text{Ra}}=2.8$, of any demonstrated clock \cite{Flambaum2009}.  The current constraint on $\dot \alpha / \alpha$ is derived from a frequency comparison between an optical clock based on the Yb$^{+}$ (E2) transition ($\tau\approx50$~ms) and a second clock based on the Yb$^{+}$ (E3) transition~\cite{Lange2021}. Considering sensitivities of demonstrated clocks, the Yb$^{+}$ (E2) transition has the second largest positive enhancement to the time variation of the fine structure constant, $\kappa_{\text{E2}}=1$, and the Yb$^{+}$ (E3) transition has largest negative enhancement, $\kappa_{\text{E3}}=-6$.  This makes the radium ion an appealing system to compare against other clocks to improve constraints on $\dot \alpha / \alpha$. 

In this Letter, we demonstrate the first operation of a radium optical clock by stabilizing a narrow-linewidth laser at 728~nm to the $7s\ ^2S_{1/2}\rightarrow$ $6d\ ^2D_{5/2}$ transition of a single $^{226}$Ra$^{+}$ ion ($I=0$).  The 728~nm laser is an external cavity diode laser stabilized to an ultralow expansion glass cavity.  We present an evaluation of the key systematic shifts and uncertainties as well as a self-referenced measurement of the clock frequency instability.  From measurements made during the clock operation, we report the first measurement of the ratio of the $D_{5/2}$ state to the $S_{1/2}$ state Land\'{e} $g$-factors.

\begin{figure}[h]
    \centering
    \includegraphics[width=\figsize\linewidth]{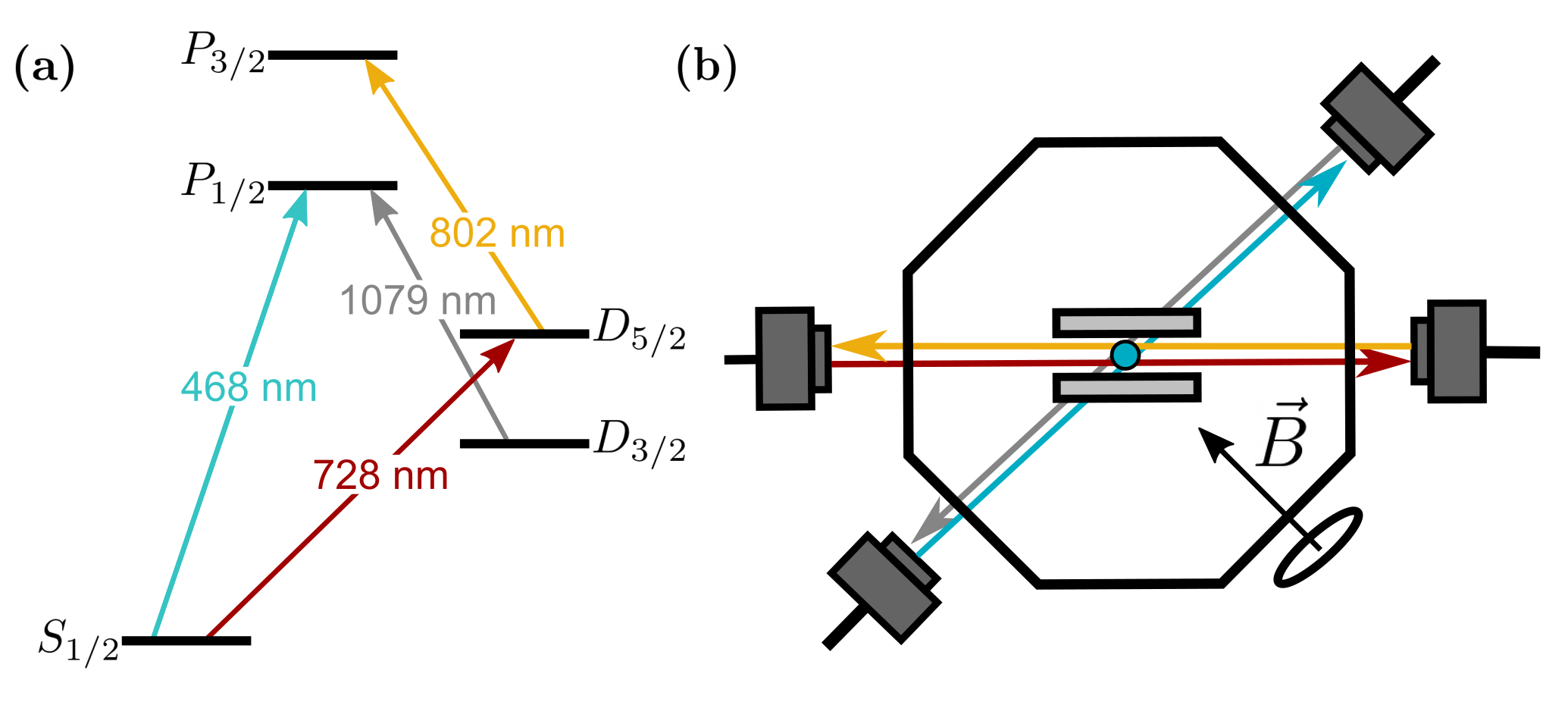}
    \caption{(a) The Ra$^{+}$ level structure for clock operation. (b) Orientation of the lasers and the magnetic field used in this Letter. The 728 nm clock beam (red) is at $45\degree$ with respect to the magnetic field to drive all ten possible Zeeman transitions between the $S_{1/2}$ and $D_{5/2}$ states.}
    \label{fig:spec}
\end{figure}

The relevant Ra$^+$ level structure, laser configuration, and quantization field used in this Letter are shown in Fig.~\ref{fig:spec}.  A single radium-226 ion is loaded by laser ablation of an $\sim$10 $\mu$Ci RaCl$_{2}$ target located 15 mm from the center of a linear Paul trap with characteristic dimensions $r_0 = 3$ and $z_0 = 7.5$~mm, see \cite{Fan2019}.  The radio frequency (rf) trap drive is operated at $\Omega_{\text{rf}}/2\pi=993$~kHz, and for a single radium ion the axial secular frequency is $\omega_{z}/2\pi=78.5$~kHz and the radial secular frequencies are $\omega_{r}/2\pi=141$ and $156$~kHz.  Acousto-optic modulators (AOMs) control the frequency and amplitude of all beams during clock operation.  Clock state readout is performed by collecting 468~nm photons scattered by the Ra$^{+}$ ion onto a photomultiplier tube \cite{Holliman2019}.  As there is no magnetic field shielding around the vacuum apparatus, each clock interrogation cycle is synchronized to the laboratory 60 Hz power line to minimize Zeeman shifts due to magnetic field fluctuations.

Linearly polarized 728~nm light is used to drive the $|S_{1/2},m=\pm1/2\rangle \rightarrow |D_{5/2},m=\pm1/2\rangle$ (C1), $|S_{1/2},m=\pm1/2\rangle \rightarrow |D_{5/2},m=\pm3/2\rangle$ (C2), and  $|S_{1/2},m=\pm1/2\rangle \rightarrow |D_{5/2},m=\pm5/2\rangle$ (C3) symmetric Zeeman transitions to operate the clock in a self-comparison mode \cite{Dube2015}.  By measuring symmetric Zeeman components that comprise all sublevels of the $D_{5/2}$ state ($|m|=1/2,3/2,5/2$), the linear Zeeman shift and the electric quadrupole shift are both canceled~\cite{Dube2013, Dube2015}.

\begin{figure}[t]
    \centering
    \includegraphics[width=\figsize\linewidth]{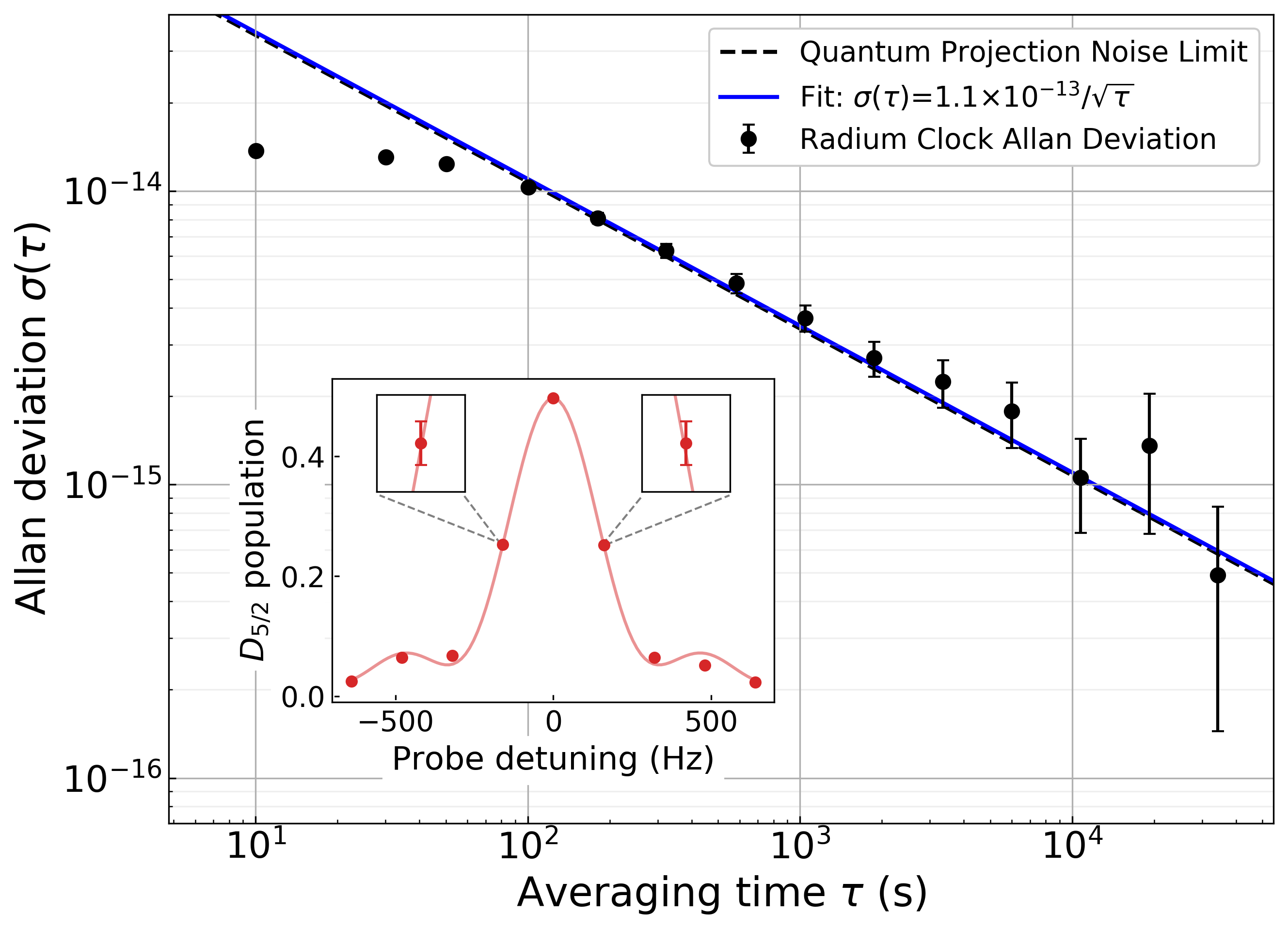}
    \caption{Allan deviation of the Ra$^{+}$ optical clock measured over $\approx$\SI{100000}{\second}.  The fractional stability (blue line) with a 3 ms interrogation time is 1.1$\times10^{-13}$/$\sqrt{\tau}$.  The quantum projection noise limit that accounts for the motional decoherence and line shape is calculated using the method described in \cite{Dube2015}.  The inset shows the $|S_{1/2},m=-1/2\rangle \rightarrow |D_{5/2},m=-1/2\rangle$ Zeeman transition averaged over the entire measurement. The HWHM lock points are magnified by $200\times$ on both axes. Error bars represent one standard deviation.  The proportion of the population driven to the excited state is limited by decoherence due to thermal motion.  The Fourier-limited linewidth of the measured transition corresponds to a 2.7 ms interrogation time.}
    \label{fig:allan}
\end{figure}

Each clock interrogation cycle begins with an initial state detection (0.5~ms) to determine correct initialization of the population into the $S_{1/2}$ or $D_{3/2}$ laser cooling states.  Following the initial state detection, the ion is Doppler cooled (5~ms) and the population is optically pumped to the appropriate $|S_{1/2},m=\pm1/2\rangle$ state (2~ms).  We then coherently interrogate the clock transition (3~ms) on either the blue- or red-detuned half width at half maximum (HWHM), after which a state detection pulse is applied.  In addition to probing the HWHM of the Zeeman transitions to determine the transition center frequency, we also interrogate the peak maximum, as well as six detunings around the peak.  For every 20 interrogation cycles of the HWHM and peak maximum, we interrogate the six detunings around the peak to ensure that symmetric Zeeman transitions are probed with equal excitation probabilities and that all transitions remain locked, see Fig. \ref{fig:allan} inset.  To reset the system, we clean out population remaining in the $D_{5/2}$ state by driving the $D_{5/2}\rightarrow P_{3/2}$ dipole transition (200~$\mu$s) where decays populate the $S_{1/2}$ and $D_{3/2}$ states.

After each interrogation cycle, the frequencies of the six Zeeman transitions are updated with individual lock servos to stabilize the clock laser's frequency to the $7s\ ^2S_{1/2}\rightarrow$ $6d\ ^2D_{5/2}$ atomic resonance. The error signal for an interrogation cycle is given by $E=(n_\text{b} - n_\text{r})/n$, where $n_\text{b}$ and $n_\text{r}$ are the number of times the population was driven to the excited state during interrogation on the blue- and red-detuned HWHM and $n=20$ is the total number of interrogations \cite{Dube2015}.  If the initial state detection determined that the population was in the $D_{5/2}$ state, the interrogation is not used in the error signal calculation.  The shift of the center detuning of each Zeeman transition is updated from the previous detuning using the error signal and the measured drift rate of the optical cavity.  Both of these values are updated throughout the experiment based on the shift of the clock transition center frequency, see Supplemental Material \cite{supplemental}. The center frequency of the $7s\ ^2S_{1/2}\rightarrow$ $6d\ ^2D_{5/2}$ transition is derived from an average of the three Zeeman pairs (C1, C2, and C3) following each interrogation cycle.  The total interrogation cycle time for the three pairs of Zeeman transitions is 10~s, where the twenty interrogation cycles of the HWHM and peak maximum takes $\sim6.1$~s, the single interrogation cycle of the six detunings around the peak takes $\sim0.6$~s, and the pulse programming and data saving takes $\sim3$~s.

The measured Ra$^{+}$ clock instability is shown in Fig.~\ref{fig:allan}.  An Allan deviation is obtained from the frequency difference of the three Zeeman pairs, (C1, C2), (C2, C3), and (C1, C3).  The average of these three Allan deviations is divided by $\sqrt{6}$ to obtain the self-referenced fractional frequency stability of ${\sigma(\tau)=1.1\times10^{-13} / \sqrt{\tau}}$, where $\tau$ is the averaging time in seconds \cite{Dube2015}. 

A summary of systematic frequency shifts and uncertainties is shown in Table \ref{table:frequency_summary}.  The overall frequency instability is currently limited by the clock interrogation time and the dead time in the total interrogation cycle.  The 3~ms interrogation time on the clock transition is mainly limited by decoherence due to short-term ambient magnetic field noise.  The 728 nm clock laser intensity used to drive a $\pi$-pulse with a $3$~ms interrogation time is 0.5(3)~kW/m$^2$, which leads to a probe-induced ac Stark shift of ${\Delta \nu / \nu = (1.7 \pm 0.9) \times 10^{-15}}$. This shift can be reduced by several orders of magnitude by implementing upgrades to the apparatus, such as magnetic field shielding and a trap that can support stronger radial confinement, which would enable interrogation times that approach the $D_{5/2}$ excited state lifetime of $\tau \sim 300$~ms \cite{Pal2009}.  Additionally, techniques such as hyper-Ramsey spectroscopy \cite{Yudin2010} or frequency stepping \cite{Taichenachev2010} could reduce this shift. All other laser beams (468, 802, 1079~nm) are turned off using double-pass AOMs during the clock interrogation pulse.  To ensure that there is no leakage light present through the AOMs, they are also backed with mechanical shutters.  During each interrogation cycle, the mechanical shutters are closed before the clock laser pulse.

Blackbody radiation generated by the finite temperature of the trapping environment causes an ac Stark shift on the clock transition which depends on the DSSP of the transition and the effective temperature of the BBR at the location of the ion.  The BBR-induced frequency shift is evaluated using the theoretical DSSP, $\Delta\alpha_0=-22.2(1.7)$ a.u. \cite{Sahoo2009a} and the effective temperature of the BBR field at the ion's location, $T_\mathrm{BBR} = 295(4)$~K \cite{Safronova2011a}.  To determine the ambient effective temperature and uncertainty observed by the ion, we measured the maximum temperature differential (3~K) of the vacuum chamber and performed a numerical simulation, using a finite element method, to estimate the maximum temperature rise (0.3~K) of the ion trap due to trap drive heating.  The resulting BBR-induced frequency shift is evaluated as ${\Delta \nu / \nu = (4.3 \pm 0.4) \times 10^{-16}}$.  At the current level of precision, the total uncertainty in the BBR shift is dominated by the uncertainty in the DSSP, and, based on previous work in Ca$^{+}$ and Sr$^{+}$, the dynamic correction to the DSSP is negligible compared to the current theoretical uncertainty \cite{Sahoo2009a, SafronovaMarc}.

During clock operation, we average the frequencies of symmetric Zeeman pairs to synthesize a clock frequency that is first-order insensitive to magnetic fields.  However, we have observed that magnetic field fluctuations at the location of the ion can be significant during the dead time between probing individual transitions in a Zeeman pair.  This effect has been observed in previous single ion clocks based on Ca$^{+}$ and can lead to a frequency shift due to a residual magnetic field drift between clock probes~\cite{Chwalla2009, Huang2012}.  The longest dead time between probings of a Zeeman pair is 50~ms, which is largely due to synchronizing the measurement with the 60 Hz ac power line.  Given an average magnetic field drift rate of ${(0 \pm 7) \times 10^{-13}}$~T/s observed in our system, and the maximum Zeeman shift sensitivity among all transitions used, $\SI{2.8e10}{}$~Hz/T, the pair averaged frequency shift is estimated to be $\Delta \nu / \nu = (0 \pm 2)\times 10^{-18}$.

Collisions between the Ra$^{+}$ ion and background gas molecules (i.e.~H$_2$) can lead to a phase shift during the clock probe pulse.  Here, we bound the corresponding clock frequency shift by assuming a worst case estimate of the phase shift of $\pm \pi/2$, which occurs in the middle of a Rabi pulse.  In this case, a collision with a background gas molecule leads to a frequency shift of 0.15$R_\text{coll}$, where $R_\text{coll}$ is the background gas collision rate~\cite{Rosenband2008}. We measure $R_\text{coll}$ in our trap to be \SI{0.0013\pm0.0004}~$\rm{s}^{-1}$ using the technique described in \cite{Fan2019a}, which corresponds to a fractional frequency shift due to background gas collisions of ${\Delta\nu / \nu = (0 \pm 6)\times 10^{-19}}$.

\begin{table}[t]
\caption{Fractional frequency shifts $(\Delta \nu / \nu)$ and uncertainties of the $^{226}$Ra$^{+}$ $7s\ ^2S_{1/2}\rightarrow$ $6d\ ^2D_{5/2}$ clock.}
\label{table:frequency_summary}
\begin{ruledtabular}
\begin{tabular}{lccc}
    Effect & Shift & Uncertainty  \\ 
    \hline \\ [-1.5ex]
    Clock laser Stark shift & $1.7\times10^{-15}$ & $9\times10^{-16}$\\
    Blackbody radiation  & $4.3\times10^{-16}$ & $4\times10^{-17}$ \\
    Magnetic field drift & 0 & $ 2\times10^{-18}$ \\
    Background gas collisions & 0 & $6\times10^{-19}$ \\
    Secular motion & $-6.0\times10^{-19}$ & $6\times10^{-19}$ \\
    Excess micromotion & $-4.2\times10^{-18}$ & $5\times10^{-19}$ \\
    Quadratic Zeeman & $4.2151\times10^{-16}$ & $1.2\times10^{-19}$ \\
    Electric quadrupole  & 0& $3\times10^{-20}$ \\ [1.ex]
   
    Total & $2.5\times10^{-15}$ & $9\times10^{-16}$
\end{tabular}
\end{ruledtabular}
\end{table}

Frequency shifts due to ion motion are characterized as that due to excess micromotion (EMM), due to the rf drive, and secular (thermal) motion.  Ion motion leads to frequency shifts due to relativistic time dilation and the ac Stark effect~\cite{Berkeland1998JAP}.  Here, the time-dilation shift is the dominant source of frequency shift and uncertainty and is expressed as ${\Delta \nu / \nu = -v^{2}/(2c^{2})}$, where $v$ is the speed of the ion in the laboratory frame and $c$ is the speed of light in vacuum.  The EMM-induced frequency shift is evaluated by measuring the amplitude of the ion motion at the trap drive frequency, $\Omega_\text{rf} / 2\pi$ \cite{supplemental}.  The frequency shift due to secular motion is evaluated by characterizing the ion temperature during clock operation~\cite{Dube2013, supplemental}.  The frequency shift due to secular motion is ${\Delta \nu / \nu = (-6.0 \pm 0.6) \times 10^{-19}}$, and the EMM-induced frequency shift is ${\Delta \nu / \nu = (-3.9 \pm 0.5) \times 10^{-18}}$.  The clock frequency shifts and uncertainties due to ion motion can be reduced by using an ion trap design that minimizes residual rf fields and supports higher secular motion frequencies.  Trap improvements and operation at the ``magic'' rf drive frequency (6.2~MHz) are expected to reduce both the magnitude and uncertainty of motional frequency shifts~\cite{Sahoo2009a}.

Additional systematic shifts, including the quadratic Zeeman shift and the electric quadrupole shift and their uncertainties are constrained at the low $10^{-19}$ level (fractional), see the Supplemental Material.

\begin{figure}
    \centering
    \includegraphics[width=\figsize\linewidth]{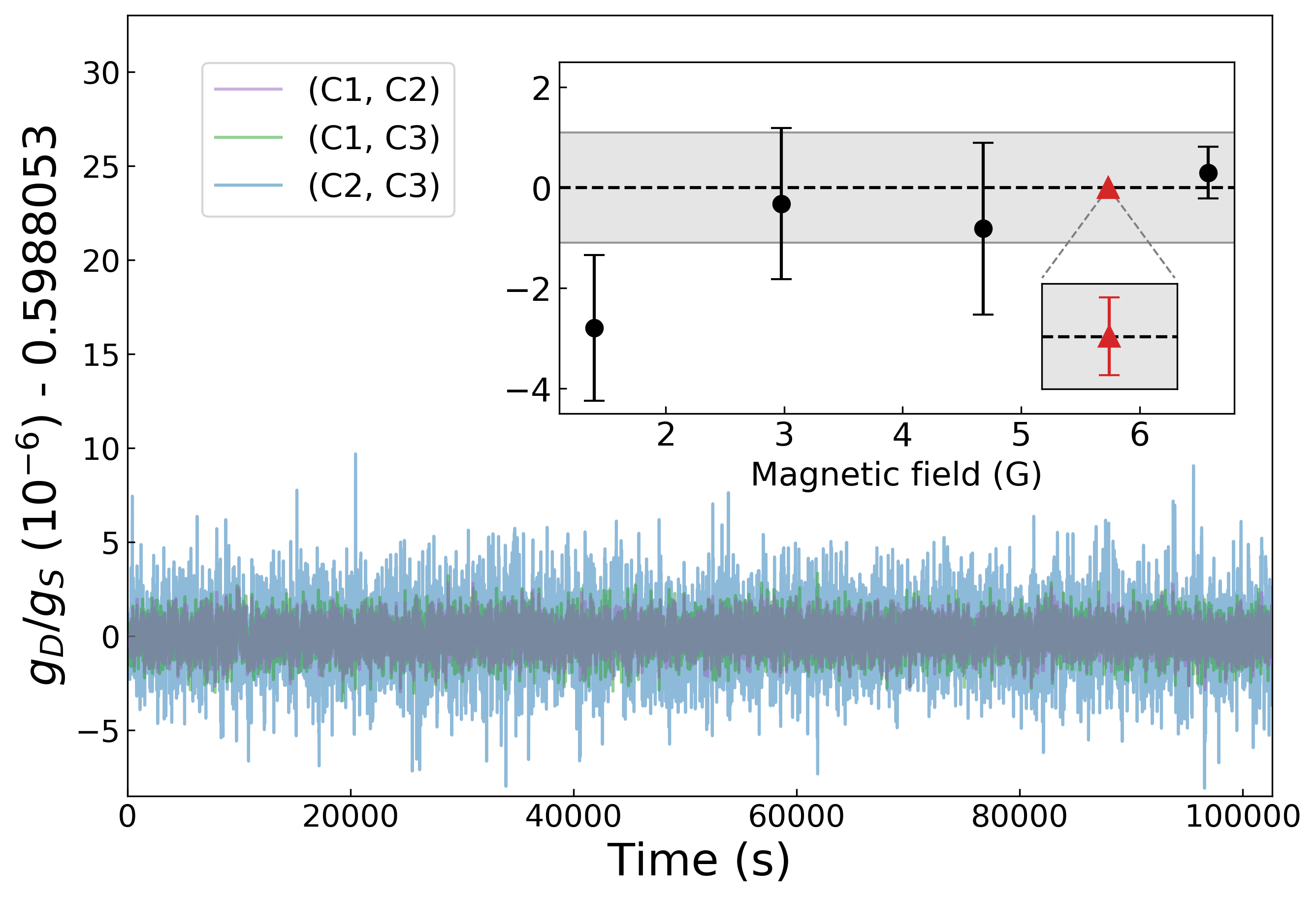}
    \caption{The three $g$-factor ratios calculated from the 3~ms interrogation time clock measurement data presented in Fig. \ref{fig:allan}.  The red triangle in the inset represents the weighted average of these data, $g_{D}/g_{S}$ = \SI{0.59880525\pm0.00000003}. The circle markers are weighted averages of $\sim$2000~s of clock data with a 250~$\mu$s interrogation time taken at a range of magnetic fields.  The shorter interrogation and measurement time results in a larger uncertainty.  The gray fill represents the full standard deviation of the five clock measurements, which we assign as the uncertainty for the $g$-factor ratio measurement.}
    \label{fig:lande_ratio}
\end{figure}

The ratio of Land\'{e} $g$-factors, $g_{D}$/$g_{S}$, is directly obtained from the clock measurement data \cite{Arnold2020}.  From a single clock measurement, such as shown in Fig. \ref{fig:allan}, we determine three ratios of Land\'{e} $g$-factors from the frequency division of the three Zeeman pairs, see Fig. \ref{fig:lande_ratio} (a).  The weighted average of these three ratios gives a value for  $g_{D}$/$g_{S}$. The reported $g_D/g_S$ ratio is calculated from a weighted average of five measurements at different magnetic fields.  The assigned uncertainty is the standard deviation of the measurements, resulting in $g_{D}/g_{S}$ = \SI{0.5988053\pm.0000011}, see Fig.~\ref{fig:lande_ratio}.  Because of the rf trapping field, an ac magnetic field is present at the trap frequency, $B_{\text{trap}}$, at the location of the ion, which shifts the measured $g_{D}$/$g_{S}$ \cite{Gan2018}.  By performing direct spectroscopy of individual Zeeman transitions with the rf trapping frequency set to the ground state magnetic sublevel splitting, we are able to set an upper bound of $B_{\text{trap}}\le7\times 10^{-8}$~T.  The systematic shift due to the maximum $B_{\text{trap}}$ value is significantly smaller than the statistical uncertainty of $g_{D}$/$g_{S}$ for all magnetic fields where the Land\'e $g$-factor ratio was measured.
To improve upon this initial measurement, the $S_{1/2}$ state Land\'{e} $g$-factor in Ra$^{+}$ could be directly measured to high precision in a Penning trap \cite{Marx1998} or in a comparison with a co-trapped ion magnetometer \cite{Rosenband2007}, which would, in turn, give the $D_{5/2}$ state Land\'{e} $g$-factor based on the ratio measured here. 

In conclusion, we have demonstrated the operation of a $^{226}$Ra$^{+}$ ion clock with a total systematic uncertainty of $\Delta \nu / \nu = 9 \times 10^{-16}$ and a frequency instability of $\sigma(\tau) = 1.1 \times 10^{-13}/\sqrt{\tau}$.  The current clock performance is primarily limited by (1) short-term magnetic field noise at the ion's location, which limits the clock interrogation time, (2) the uncertainty in the DSSP that dominates the uncertainty in the frequency shift due to BBR, and (3) limitations in the trap design that lead to motional decoherence.  The ambient magnetic field noise can be reduced by adding magnetic field shielding, as has been done with $^{40}$Ca$^{+}$ \cite{Chwalla2009} and $^{88}$Sr$^{+}$ \cite{Dube2013} and motional decoherence can be reduced by using an improved trap design~\cite{Brewer2019, Clements2020}.  Reduced magnetic field sensitivity could be realized with radium-225, which has first-order magnetic field insensitive states due to its $I=1/2$ nuclear spin.  The 14.9 day half-life of radium-225 can be overcome by using an oven based on the decay of thorium-229 ($\tau_{1/2}\approx7340$~y), as demonstrated with a 10~$\mu$Ci oven source \cite{Santra2014}.  Such a source promises a long-term \emph{in vacuo} supply of radium-225, as the thorium vapor pressure is more than a trillion times smaller than radium \cite{crc}, which also makes it robust to inadvertently exhausting the atom supply by running the oven at high temperatures \cite{Santra2013}.  These features, along with the photonic-technology compatible wavelengths of Ra$^{+}$ and the low optical power requirements of an ion clock make it an intriguing candidate for a transportable optical clock.  

We thank Pierre Dub\'e, Shimon Kolkowitz, Galan Moody, and Jwo-Sy Chen for helpful discussions and feedback.  We thank Mingda Li for finite element analysis simulations and Marianna Safronova for sharing calculated values of the dynamic polarizability. This research was performed under the sponsorship of the ONR Grant No. N00014-21-1-2597, NSF Grant No.~PHY-1912665, NSF Grant No.~PHY-2110102, and the University of California Office of the President Grant No. MRP-19-601445.


%

\end{document}


\title{Supplemental Material}

\maketitle

\renewcommand\thefigure{S\arabic{figure}} 
\renewcommand\theequation{S\arabic{equation}} 

\section{Clock Measurement Sequence}

State Detection: The signal for each interrogation are the 468 nm photons spontaneously emitted by the radium ion, which are collected onto a photomultipler tube (PMT), Hamamatsu H10682-210. During 0.5~ms of state detection 25 photons are collected on average if the population is in the $S_{1/2}$ or $D_{3/2}$ states, and 1.5 photons of background scattered light are collected if the population is shelved in the $D_{5/2}$ clock state. We set the bright-state detection threshold to 5.5 counts.

State Preparation: We simultaneously drive the $|S_{1/2},m=\mp1/2\rangle \rightarrow |D_{5/2},m=\pm3/2\rangle$, $D_{3/2}\rightarrow P_{1/2}$, and $D_{5/2}\rightarrow P_{3/2}$ transitions to prepare the population in the $|S_{1/2},m=\pm1/2\rangle$ state. 

Cleanout: To prepare for the next interrogation any population shelved in the clock state is cleaned out with light at 802 nm, where decays from the $P_{3/2}$ state populate the $S_{1/2}$ and $D_{3/2}$ states.

\begin{figure*}[h]
    \centering
    \includegraphics[width=0.8\linewidth]{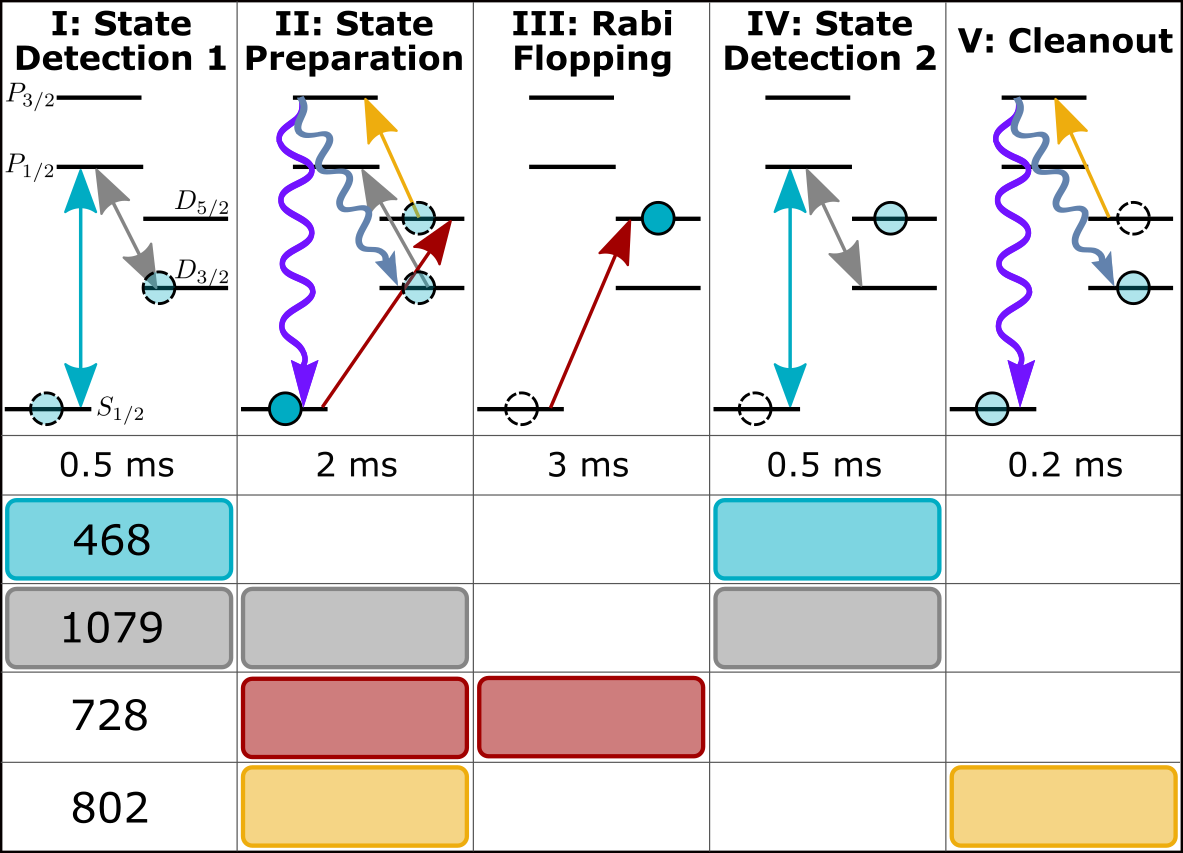}
    \caption{The measurement sequence for a single interrogation. There is a 5 ms Doppler cool period after the first state detection which is not shown in the figure. Squiggly lines depict E1 allowed decays, straight lines show optical pumping transitions, and double-arrows indicate optical cycling transitions.}
    \label{fig:branching_ratio}
\end{figure*}

\section{Lock Servo}
The individual locking servos update the frequency detuning of each Zeeman transitions by $\Delta\nu$, following,

\begin{equation}
    \Delta\nu = GE + r_\text{cavity}t_c,
\end{equation}

\noindent where $E=(n_b - n_r)/n$ is the error signal defined in the text, $G$ is a gain parameter, $t_c$ is the total interrogation cycle time (10~s), $r_\text{cavity}$ is the cavity drift rate ($\sim$0.2~Hz/s). The cavity drift rate is updated with the average frequency of all transitions probed using an integral servo, with a time constant of $\sim$1000~s. 

\section{Clock Systematics}

\subsection{Excess micromotion shifts}

\subsubsection{Second-order Doppler shift}

The second-order Doppler shift in fractional form is 

\begin{equation}
    \frac{\Delta\nu_\text{D2}}{\nu_0} = -\left(\frac{\Omega}{\omega_0}\right)^2\sum_{x,y,z} R_i,
\end{equation}
where $\nu_0$ is the radium clock transition frequency in Hz, $\Omega$ is the trap rf drive frequency, $\omega_0=2\pi\nu_0$, and $\sum R_i$ is the sum of the micromotion sideband intensity ratios \cite{Dube2013}.  We use $\nu_0=\SI{412007.701\pm0.018}{\giga\hertz}$ \cite{Holliman2019}, $\Omega_\text{rf}/2\pi=993~$kHz, and $\sum R_i=0.74(8)$, which we determine from measuring the micromotion sidebands in three near-orthogonal directions.  We calculate $\Delta\nu_\text{D2}=-1.77(19)$~mHz.

\subsubsection{Scalar Stark shift}

The scalar Stark shift in fractional form is 

\begin{equation}
    \frac{\Delta\nu_\text{scalar}}{\nu_0} = -\frac{\Delta\alpha_0}{\hbar\omega_0}\left(\frac{m\Omega^2 c}{e\omega_0}\right)^2\sum_{x,y,z} R_i,
\end{equation}
where $\Delta\alpha_0$ is the differential static scalar polarizability and $m$ is the mass of the ion \cite{Dube2013}. We use $\Delta\alpha_0=-22.2(1.7)$ a.u. \cite{Sahoo2009a}.  We calculate $\Delta\nu_\text{scalar}=0.045(6)$~mHz.

\subsubsection{Tensor Stark shift}

The maximum possible tensor stark shift in fractional form is

\begin{equation}
    \frac{\Delta\nu_\text{tensor}}{\nu_0} = -\frac{\alpha_2}{\hbar\omega_0}\left(\frac{m\Omega^2 c}{e\omega_0}\right)^2\sum_{x,y,z} R_i,
\end{equation}
where $\alpha_2$ is the $D_{5/2}$ state tensor polarizability \cite{Dube2013}. We use $\alpha_2(D_{5/2})=-52.6(5)$ a.u. \cite{Sahoo2009a}.  We calculate $\Delta\nu_\text{tensor}=0.108(12)$~mHz. We use the sum of the maximum shift and uncertainty, 0.120~mHz, as the uncertainty for this shift. 

\subsection{Thermal motion shift}

\subsubsection{Second-order Doppler shift}
The fractional second-order Doppler shift due to the ion's secular motion is

\begin{equation}
    \frac{\Delta\nu_{\text{D2}}}{\nu_0} = -\frac{5kT_\text{ion}}{2mc^2},
\end{equation}
where $T_\text{ion}$ is the temperature of the ion, and $k$ is Boltzmann's constant \cite{Berkeland1998JAP, Dube2013, Keller2015}.  We measured the ion temperature along in the axial and radial mode directions by driving Rabi oscillations on the clock transition and determined an average temperature  of $T_\text{ion} = 0.6$~mK. We estimate the temperature uncertainty to be 0.6~mK to account for fluctuations in Doppler cooling laser frequencies and any temperature imbalance between different secular motion modes. This measured $T_\text{ion}$ is consistent with the Doppler cooling limit $T_c=0.4$~mK.  The corresponding frequency shift is $\Delta\nu_{\text{D2}}=-0.3(3)$~mHz. 

\subsubsection{Scalar Stark shift}

The fractional scalar Stark shift due to the ion's thermal motion is

\begin{equation}
    \frac{\Delta\nu_{\text{scalar}}}{\nu_0} = -\frac{kT_\text{ion}}{mc^2}\frac{\Delta\alpha_0}{\hbar\omega_0}\left(\frac{m\Omega c}{e}\right)^2,
\end{equation}
where $T_\text{ion}$ is the temperature of the ion \cite{Berkeland1998JAP, Dube2013, Keller2015}.  Here, the shift is measured to be $\Delta\nu_{\text{scalar}}=0.003(3)$~mHz.

\subsection{Electric quadrupole shift}

The electric quadrupole shift is cancelled because we are averaging Zeeman pairs. We can approximate the uncertainty on this cancellation from the electric quadrupole shift magnitude, $A_Q=[\Delta f_Q(m_{j'}=5/2)-\Delta f_Q(m_{j'}=1/2)]/6$ \cite{Dube2013}.  From the frequency difference between shifted pairs we determine the average drift rate of $A_Q$ to be $-1.3~\mu$Hz/s.  Given our cycle time of 10~s we estimate the electric quadrupole shift uncertainty to be 0.013 mHz. 

\subsection{Clock laser ac Stark shift}

\textbf{728 nm:}  The ac Stark shift due to off-resonant dipole coupling is

\begin{equation}
    \Delta\nu_\text{ac}(\lambda)=-\frac{\Delta\alpha_\text{ac}(\lambda)}{2hc\epsilon_0}I=\kappa(\lambda)I,
\end{equation}
where $\Delta\alpha_\text{ac}(\lambda)=\Delta\alpha_\text{ac}(D_{5/2},\lambda)-\Delta\alpha_\text{ac}(S_{1/2},\lambda)$, $\lambda$ is the laser wavelength and $\kappa(\lambda)$ is the ac Stark shift intensity coefficient \cite{Dube2013}. Using  $\Delta\alpha_\text{ac}(D_{5/2}, \text{728~nm})=-169(4)$~a.u. and $\Delta\alpha_\text{ac}(S_{1/2},\text{728~nm})=140(3)$~a.u. \cite{marianna2021}, we calculate $\kappa(\text{728~nm})=1.45(2)$~mHz/(W/m$^2$).  We measure the clock laser intensity to be 500(300)~W/m$^2$.  The large uncertainty in the clock laser intensity is due to probing all Zeeman transitions with equal contrast, which requires different AOM driving amplitudes.    We calculate the clock laser ac Stark shift to be 0.7(4)~Hz.  

\subsection{Quadratic Zeeman shift}

The quadratic Zeeman shift is caused by the mixing of the $D_{5/2}$ and $D_{3/2}$ sublevels when the ion is exposed to a magnetic field. This shift is

\begin{equation}
    \nu_{ZQ} = \langle\nu'_{ZQ}(m_j)\rangle=\frac{2}{15}\left(\frac{[\mu_B B(g_s-1)]^2}{h^2\nu_{DD}}\right),
\end{equation}
where $B$ is the magnetic field, $g_S=-g_e$, $\mu_B$ is the Bohr magneton, and $\nu_{DD}$ is the energy separation between the $D_{5/2}$ and $D_{3/2}$ states \cite{Dube2013}.  We use a static magnetic field of $B=0.00057369(8)$~T, and $\nu_{DD}=\SI{49.73002\pm0.00005}{\tera\hertz}$ \cite{Holliman2019}.  We calculate $\nu_{ZQ}=173.66(5)$~mHz. 

\section{Evaluation of the Land\'e g-factor Ratio}

From the clock measurement data we calculate three Land\'e $g$-factor ratios from the three Zeeman pairs (C1, C2, and C3) defined in the text. For a given Zeeman transition the frequency detuning from the 728 nm transition center is given by 

\begin{equation}
    \delta = \kappa\frac{g_{D}}{g_{S}}m_{D} - \kappa m_{S},
\end{equation}

\noindent where $\kappa=g_{S} \mu_{\text{B}} B/h$, and $g_{S}$ is the ground state Land\'e $g$-factor, $\mu_{\text{B}}$ is the Bohr magneton, $B$ is the magnetic field, and $h$ is the Planck constant. The C1, C2, and C3 frequency detunings can be written as
\begin{align*}
    \text{C1} &= \delta_{1/2\rightarrow 1/2} - \delta_{-1/2\rightarrow -1/2} = \kappa\left(\frac{g_{D}}{g_{S}} - 1\right)\\
    \text{C2} &= \delta_{1/2\rightarrow 3/2} - \delta_{-1/2\rightarrow -3/2} = \kappa\left(3\frac{g_{D}}{g_{S}} - 1\right)\\
    \text{C3} &= \delta_{1/2\rightarrow 5/2} - \delta_{-1/2\rightarrow -5/2} = \kappa\left(5\frac{g_{D}}{g_{S}} - 1\right)
\end{align*}

\noindent The frequency ratios of C1, C2, and C3 then directly give three Land\'e $g$-factor ratios

\begin{equation}
    \frac{ g_{D}}{g_{S}} = \frac{\frac{C1}{C3} - 1}{5\frac{C1}{C3}-1} = \frac{\frac{C1}{C2} - 1}{3\frac{C1}{C2}-1} = \frac{\frac{C2}{C3} - 1}{5\frac{C2}{C3}-3}
\end{equation}

\noindent This method cancels out the electric quadrupole shift as C1, C2, and C3 are the frequency differences between symmetric Zeeman transitions.

\section{Autler Townes Shift}
When the energy splitting between the ground state Zeeman sublevels, $\omega_S = g_S\mu_{\text{B}}B_0/\hbar$, equals the rf trapping frequency, $\Omega_\text{rf}/2\pi=993~$kHz, there is an Autler-Townes splitting ($g_{S} \mu_{\text{B}} B_\text{trap}/2\hbar$) of the clock transition due to the trap's magnetic field, $B_\text{trap}$, driving the $|S_{1/2},m=\pm1/2\rangle \rightarrow |S_{1/2},m=\mp1/2\rangle$ transition. To bound $B_\text{trap}$, we set $\Omega_\text{rf}=\omega_{S}$ within 1~kHz by changing the static magnetic field $B_{0}$.  We did not observe any splitting at the sub 200~Hz level during direct spectroscopy of individual Zeeman transitions, which indicates that the Rabi frequency due to the trap's magnetic field is not larger than $\sim$1 kHz.  Therefore the corresponding $B_\text{trap}$ must be no larger than $7\times10^{-8}$ T.  The $B_{\text{trap}}$ shifts the Land\'e $g$-factor ratio, $g_D/g_S$, by 

\begin{equation}
\frac{g_D}{g_S} = \left[\frac{1+\frac{1}{2}\frac{\omega_D^2}{\omega_D^2-\Omega_\text{rf}^2}\frac{\langle B_\text{trap}^2\rangle}{B_0^2}}{1+\frac{1}{2}\frac{\omega_S^2}{\omega_S^2-\Omega_\text{rf}^2}\frac{\langle B_\text{trap}^2\rangle}{B_0^2}}\right]\text{r},
\end{equation}

\noindent where  $\omega_D = g_D\mu_{\text{B}}B_0/\hbar$ and r is the measured Land\'e $g$-factor ratio at the given static magnetic field \cite{Arnold2020}. The shift in the measured Land\'e $g$-factor ratio due to $B_{\text{trap}}$ is much smaller than the statistical uncertainty of the measured Land\'e $g$-factor ratio.


%